\documentclass[11pt,B5]{article}
\usepackage{graphicx}

\begin{document}
\title {Strangeness, Cosmological Cold Dark Matter and Dark Energy}
\date{}
\maketitle 
%\vspace{-1in}
\author{\begin{center}{\underline{Sibaji Raha}$^1$
\footnote{Electronic Mail :
sibaji@bosemain.boseinst.ac.in},
Shibaji Banerjee$^2$, 
Abhijit Bhattacharyya$^3$,
Sanjay K. Ghosh$^1$,
Ernst-Michael Ilgenfritz$^4$, 
Bikash Sinha$^5$, Eiichi Takasugi$^6$ and Hiroshi Toki$^4$}
\end{center}
$^1$Physics Department, Bose Institute, 93/1, A. P. C. Road, 
Kolkata 700009, INDIA \\
$^2$Physics Department, St. Xavier's College, 
30, Park Street, Kolkata 700016, INDIA \\
$^3$Physics Department, Scottish Church College, 1 \& 3, Urquhart Square, Kolkata 700006, INDIA \\
$^4$Research Center for Nuclear Physics, Osaka University, Ibaraki, 
Osaka 567-0047, JAPAN \\
$^5$ Saha Institute of Nuclear Physics, 1/AF, Bidhannagar, Kolkata - 700 064, INDIA\\
$^6$ Graduate School of Physics, Osaka University, Toyonaka, Osaka 560-0043, JAPAN}

\begin{abstract}
It is now believed that the universe is composed of a small amount of the normal luminous matter, a substantial amount of matter (Cold Dark Matter: CDM) which is non-luminous and a large amount of smooth energy (Dark Energy: DE). Both CDM and DE seem to require ideas beyond the standard model of particle interactions. In this work, we argue that CDM and DE can arise entirely from the standard principles of strong interaction physics out of the same mechanism.
\end{abstract}

\vspace{1in}

The current consensus \cite{1,2} in cosmology is that the standard 
model comprising an initial Big Bang and a flat universe accommodates only about 4\% atoms and \( \sim\) 23\% cold dark matter (CDM). The remaining \(\sim\) 73\% is a smooth energy, called the dark energy (DE). Matter contributing to CDM should have a dust-like equation of state, pressure \( p \approx \) 0 and energy density \( \rho > 0 \) and be responsible for clustering on galactic or supergalactic scales. Dark energy (DE), on the other hand, shows no clustering features at any scale. It is required to have an equation of state \( p = w\rho \) where \( w< 0 \) (ideally \( w = -1 \)), so that for a positive amount of dark energy, the resulting negative pressure would facilitate an accelerated expansion of the universe, evidence for which has recently become available from the redshift studies of type IA supernovae \cite{3}. The present-day crtitcal density is \(\sim 10^{-47}\) GeV\(^4\) so that \( \rho_{DE} \) today is \(\sim 10^{-48} \) GeV\(^4\).

There is no agreement within the community about the origin or the nature of CDM or DE. It is argued that given the limit of \( \Omega_B \sim 0.04 \), CDM cannot be baryonic; as a result, various exotic possibilities, all beyond the standard model \( SU(3)_c \times SU(2) \times U(1) \) of particle interactions, have been suggested. The situation is even more complicated for DE. The most natural explanation for DE would be a vacuum energy density, which {\it a priori} would have the correct equation of state (\( w = -1 \)). This possibility however is beset with the insurmountable difficulty that for any known (or conjectured) type of particle interaction, the vacuum energy density scale turns out to be many orders of magnitude larger than the present critical density. A trivial, but aesthetically displeasing, way out could be to {\it a priori} postulate a small cosmological constant. There exists in the recent literature a large number, too many to cite here, of speculative suggestions, all substantially beyond the standard model. Apart from being non-standard, all these pictures require large amounts of fine-tuning. In this work, we argue that there is no essential need to go beyond the standard model \( SU(3)_c \times SU(2) \times U(1) \) to understand the nature of CDM and DE; they can both arise from the same process of the cosmic quark-hadron phase transition occurring during the microsecond epoch after the Big Bang. 

The role of phase transitions \cite{4} in the early universe has been recognized to be of paramount importance. In the strong interaction (Quantum Chromodynamics) sector, there is expected to be a phase transition separating the confined (hadronic) phase from the deconfined (quark-gluon plasma) phase. In the early universe, this phase transition is predicted to occur during the microsecond epoch after the Big Bang. The order of this phase transition is at present an open issue. While it may be of second order (or even a cross-over transition) in the laboratory, the cosmic phase transition could most likely be of first order, as has been argued earlier\cite{5,6,7}. In what follows, we shall tacitly assume it to be of first order; the conclusions may however be valid even if the transition is not strictly of first order. We shall return to this issue at the end of the discourse. Another crucial {\it ansatz} in our scenario is that the universe is overall colour neutral at all times. We further assume, in keeping with the standard cosmological model, that the baryon number of the universe has been generated much before the universe reaches the microsecond era. The net baryon number till this epoch is carried in the form of (net) quarks.

A first order phase transition can be described through a bubble nucleation scenario. At temperatures higher than the critical temperature \( T_c \), the coloured quarks and gluons are in a thermally equilibrated state in the perturbative vacuum (the quark-gluon plasma or QGP). The total colour of the universe is neutral so that the total colour wave function of the universe is a singlet. Then, as \( T_c \) is reached and the phase transition starts, bubbles of the hadronic phase begin to appear in the quark-gluon plasma, grow in size and form an infinite chain of connected bubbles (the percolation process). At this stage, the ambient universe turns over to the hadronic phase. Within this hadronic phase, the remaining high temperature quark phase gets trapped in large bubbles. As is well known, this process is associated with a fluctuation in the temperature around \(T_c\); the bubbles of the hadronic phase can nucleate only when the temperature falls slightly below \(T_c\). The released latent heat raises the temperature again and so on. It is thus fair to assume that the temperature of the universe remains around \(T_c\) at least upto percolation.

Witten \cite{6} argued some time ago that the net baryon number contained in these Trapped False Vacuum Domains (TFVD) could be many orders of magnitude larger than that in the normal hadronic phase and they could constitute the absolute ground state of strongly interacting matter. It has been shown \cite{8} that if these TFVDs possess baryon number in excess of \( 10^{42-44} \), they would be stable on cosmological time scales and would form the so-called strange quark nuggets (SQN). (It should be mentioned here that the role of strangeness is extremely important in this context; it is the population of the strange quark sector through weak interaction which ensures the stability of the SQNs against decay into normal hadrons. Hence the occurrence of the term "Strangeness" in the title and the justification of the topic in the present conference.) In such situations, they would have spatial radii $\sim$ 1 m, while their spatial separation would be $\sim$ 300 m \cite{9}. TFVDs with baryon number less than this would evaporate into normal baryons quite rapidly, much before Big Bang Nucleosynthesis (BBN) \cite{10} starts. (To distinguish the baryon number contained in SQNs from that participating in BBN, we denote the SQN matter as quasibaryonic.) The SQNs could evolve \cite{9} to primordial structures of approximate solar mass, which could manifest themselves as the Massive Compact Halo Objects (MACHO), discovered \cite{11,12} through gravitational microlensing in the Milky Way Halo in the direction of the Large Magellanic Cloud (LMC).

In all these considerations, the explicit role of colour, the fundamental charge of strong interaction physics, has been glossed over. It has been tacitly assumed that in a many-body system of quarks and gluons, colour is averaged over, leaving only a statistical degeneracy factor for thermodynamic quantities. We argue that such simplification has led us to overlook a fundamentally important aspect of strong interaction physics in cosmology. 

Let us now understand the QGP characteristics in terms of the Debye screening length (DSL). All colour charges are neutralised within the DSL (\( \sim  \frac{1}{g_s(T) T} \), \( g_s \) being the strong coupling conatant). Formation of hadrons becomes possible only when the DSL becomes larger than the typical hadronic radius. The existence of QGP would mean that there are sufficient number of colour charges present within the Debye volume. It can be shown that upto \( T_c \), Debye length is less than a fermi and more than 10 colour charges are present within the Debye volume (quarks, antiquarks and gluons).  On the other hand, the net number of quarks (obtained using \( \frac{n_b}{n_\gamma} \sim 10^{-10} \) ) is much less than one within the same volume. Thus, to ensure both colour neutrality as well as integer baryon number,  one would need a long-range correlation beyond the Debye length in QGP, the quantum entanglement property \cite{13}.

We will now consider the process of the cosmic quark-hadron phase transition from the quantum mechanical standpoint of colour confinement. As already mentioned, the colour wave function of the entire universe prior to the phase transition must be a singlet (this assumption is at the same level as that of total electric charge of the universe being zero), which means it cannot be factorized into constituent product states; the wave functions of all coloured objects are completely entangled \cite{13} in a quantum mechanical sense. In such a situation, the so-called quark-gluon plasma phase, the universe is characterized by a vacuum energy corresponding to the perturbative vacuum of Quantum Chromodynamics (QCD). As the phase transition proceeds, locally colour neutral configurations (hadrons) arise, resulting in gradual decoherence of the entangled colour wave function of the entire universe. Note that the coloured objects within the hadrons are entangled among themselves but not with those in the rest of the universe. This amounts to a proportionate reduction in the perturbative vacuum energy density, which goes into providing the latent heat of the transition, as well as the mass and the kinetic energy of the particles in the non-perturbative (hadronic) phase. (It should be mentioned here that the vacuum energy of the non-perturbative phase of QCD is taken to be zero; more on this later.) In the quantum mechanical sense of entangled wave functions, the end of the quark-hadron transition would correspond to complete decoherence of the colour wave function of the universe; the entire vacuum energy would disappear as the perturbative vacuum would be replaced by the non-perturbative vacuum. 

Combining these observations with the formation of TFVDs as discussed above, it is obvious that in order for the TFVDs to be stable physical objects, they must be colour neutral. This is synonymous with the requirement that they all have integer baryon numbers, i.e., at the moment of formation each TFVD has net quark numbers in exact multiples of 3. For a statistical process, this is, obviously, most unlikely and consequently, most of the TFVDs would have some residual colour at the percolation time. Then, on the way to becoming colour singlet, they would each have to shed one or two coloured quarks. This scenario is certainly not inconsistent with the screening of colour within DSL, even if the size of the TFVD is much larger than the DSL. It is well known from the Electromagnetic plasma that the local charge neutrality is violated over a small length of the order of DSL at the boundary of the plasma (plasma sheath effect).

Thus, the end of the cosmic QCD phase transition corresponds to a situation where there would be a few quarks, separated by spacelike distances. It has to be noted that such a large separation, apparently against the dictates of QCD, is by no means unphysical. The separation of coloured TFVDs occurs at the temperature \( T_c \), when the effective string tension is zero, so that there does not exist any long range force. Even more importantly, these orphan quarks are not deconfined at all; they do not form asymptotic states. In terms of the quantum entanglement and decoherence of the colour wave function, their colour wave functions must still remain entangled and a corresponding amount of the perturbative vacuum energy would persist in the universe. In a physical picture, the orphan quarks, being unable to form strings and recombine into baryons, belong to a very dilute many-body system of quarks confined in a very large bag which spans the entire universe.  

Note that the above scenario is unique for the early universe. For the QGP putatively formed in the laboratory in energetic heavy ion collisions, the  process is limited to strong interaction time scales so that the size of the system is of the order of a few fermis, comparable to the DSL. Thus separation of quarks over large spatial distances is not at all likely. Furthermore, one has to also take into account the two possible situations in these collisions. If there is complete or substantial stopping, as the case seems to be till present energies, the baryon number density is very high so that there would be sufficient number of net quarks within the Debye volume. On the other hand, if there is total transparency, the baryon chemical potential in the central region would be zero. In either case, there is no {\it a priori} need to invoke quantum entanglement.
 
There does not exist any way to calculate the perturbative vacuum energy from first principles in QCD. For the latter quantity, one may adopt the phenomenological Bag model \cite{14} of confinement, where the Bag parameter B (\(\sim (145 MeV)^4 \)) is the measure of the difference between the perturbative and the non-perturbative vacua. Thus we can assume that at the beginning of the phase transition, the universe starts out with a vacuum energy density B, which gradually decreases with increasing decoherence of the entangled colour wave function. A natural thermodynamic measure of the amount of entanglement during the phase transition could be the volume fraction (\( f_q \equiv _{colour}/V_{total} \)) of the coloured degrees of freedom; at the beginning, \(f_q\) is unity, indicating complete entanglement, while at the end, very small but finite entanglement corresponds to a tiny but non-zero \(f_q\) due to the coloured quarks. Accordingly, the amount of perturbative vacuum energy density in the universe at any time is the energy density B times the instantaneous value of \(f_q\); within the scenario discussed above, the remnant perturbative vacuum energy at the end of the QCD transition would just be B \( \times f_{q,O}\), where \(f_{q,O}\) is due solely to the orphan quarks. An order of magnitude estimate for \(f_{q,O}\) can be carried out in the following straightforward manner. On the average, each TFVD is associated with 1 orphan quark so that the number \( N_{q,O} \) of orphan quarks within the horizon volume at any time is about the same as the number \( N_{TFVD} \) of TFVDs therein. It is well known from the study of percolating systems \cite{15} that percolation is characterized by a critical volume fraction \(f_c \sim \) 0.3 of the high temperature phase. In the present case, this would require \(f_q \) in the form of TFVD-s to be \( \sim \) 0.3. Following the ansatz of Witten \cite{6} that the most likely length scale for a TFVD is a few cm, one can estimate \(N_{TFVD}\) (and hence \(N_{q,O}\)) within the horizon at the percolation time of about 100 \(\mu\)sec \cite{5} to be about 10\(^{18-20}\). The inter-TFVD separation comes out to be \( \sim \) 0.01 cm at that time. (It is obvious that the orphan quarks, separated by distances of 0.01 cm, cannot develop colour strings between them, even if there is some non-zero string tension generated at temperatures slightly lower than \(T_c\).) Then, if we naively associate an {\it effective} radius of \(\sim 10^{-14} cm\) (estimated from \( \sigma_{qq} = \frac{1}{9} \sigma_{pp}~;~~ \sigma_{pp} \sim 20 mb \) ) with each orphan quark, we obtain \( f_{q,O} \sim N_{q,O} \times (v_{q,O}/V_{total}) \sim 10^{-42} - 10^{-44}\) (where $v_{q,O}$ is the effective volume of an orphan quark), so that the residual pQCD vacuum energy comes out to be in the range 10\(^{-46}\) to 10\(^{-48} GeV^4\), just the amount of DE. 

Even though the DE component appears during the microsecond era in the history of the universe, it remains negligible in comparison to the matter density for most of the history. Since matter density decreases as \( R^{-3} \) (R being the scale size), while DE density remains constant, the latter can become dominant only at very late times (\( z \sim 0.17 \)) and thus would not affect the galaxy formation scenarios to any extent.

The density of the orphan quarks in the present universe is exceedingly small; compared to \( \sim 10^{77} \) baryons which took part in BBN, there would be \( 10^{44-45} \) orphan quarks. Their flux in the cosmic rays would be negligibly small; non-observance of fractionally charged objects is thus not a detractor.

The scenario presented here may still remain valid even if the QCD phase transtition is not strictly of first order, provided there are finite size fluctuation associated with the quark-hadron transition. A definitive answer to this question would require a detailed simulation which is a cherished goal.

{\it A posteriori}, it is tempting to mention that it should perhaps have been anticipated that if the cosmological constant does arise from a vacuum energy, then the QCD vacuum is the most natural candidate. It is a known (and accepted) lore of (renormalisable) quantum field theories that the divergent vacuum energy density is renormalised to the physical parameters of the theory. Within all the field theories now in vogue, it is only the QCD which has a distinct perturbative and a non-perturbative vacuum, which are separated by a finite energy density, irrespective of the renormalisation prescription. (That we are unable to estimate this as yet from first principles in QCD is a technical shortcoming, not a conceptual one.) It is thus most plausible that even if suitable renormalisation prescriptions remove all vacuum energy densities, the finite part of the perturbative vacuum energy density of QCD should survive and play the role of the cosmological constant.

We therefore conclude by reiterating that the emergence of both CDM and DE from the same mechanism {\it entirely within the standard model of particle interactions} is a very interesting possibility which deserves detailed attention.

A natural corollary of the present scenario would be the existence of strange quark matter (strangelets) in the cosmic ray flux. Despite many searches, no conclusive evidence for the existence of strangelets has yet emerged. The reader is referred to the article of Jes Madsen in this volume for a review of the status of such searches. We are in the process of setting up a dedicated large area array of passive detectors at mountain altitudes in Eastern Himalayas for such detection, funding for which has very recently been approved. 

SR would like to thank the Research Center for Nuclear Physics, Osaka University for their warm hospitality during his sojourn there, where this work was initiated.

\end{document}